\documentclass[aps,amsmath,amssymb,twocolumn,showpacs,prl]{revtex4}
\usepackage{graphicx,color}
 \graphicspath{{./}{./figures/}}
\usepackage{dcolumn}
\usepackage{bm}

\definecolor{labelkey}{cmyk}{.4,.2,0,0}

\newcommand{\titre}[1]{}

\newcommand{\beq}{\begin{equation}}
\newcommand{\eeq}{\end{equation}}
\newcommand{\be}{\begin{equation}}
\newcommand{\ee}{\end{equation}}
\newcommand{\bea}{\begin{eqnarray}}
\newcommand{\eea}{\end{eqnarray}}

\newcommand{\rmd}{\mathrm{d}}

\begin{document}

\title{Shock statistics in higher-dimensional Burgers turbulence}
\author{Pierre Le Doussal$^1$, Alberto Rosso$^2$, and Kay J\"org Wiese$^1$}
 \affiliation{${}^1$CNRS-Laboratoire
de Physique Th{\'e}orique de l'Ecole Normale Sup{\'e}rieure, 24 rue
Lhomond,75005, Paris, France.\\
$^{2}$CNRS-Laboratoire de Physique Th\'eorique et Mod\`eles Statistiques, 
Universit\'e Paris-Sud, 91405 Orsay, France.
} 

\begin{abstract}
We conjecture the exact shock statistics in the inviscid decaying Burgers
equation in $D>1$ dimensions, with a special class of correlated initial velocities, which reduce 
to Brownian for $D=1$. The prediction is based on a field-theory argument, and receives support
from our numerical calculations. We find that, along  any
given direction, shocks sizes and locations are uncorrelated. 
\end{abstract}


\maketitle

Decaying turbulence is  characterized by the existence of an inertial range  in the inviscid limit (small
viscosity limit) with scaling and multi-scaling. These features are shared by simpler models as passive advection \cite{GawedzkiKupiainen1995
} and the Burgers equation \cite{Burgers74} where the velocity field present manifolds of discontinuities called {\it shocks}.  How the velocity field evolves 
from a prescribed random initial condition $\vec v(\vec r,t=0)$, $\vec r \in \mathbb{R}^D$,
and the statistics of these shocks, are of high interest. 
Unfortunately,  exact results are restricted to  few solvable cases in space dimension 
$D=1$ \cite{Burgers74,Kida1979,Sinai1992,SheAurellFrisch1992,Bertoin1998,FrachebourgMartin2000,LeDoussal2006b,LeDoussal2008,Valageas2009,FyodorovLeDoussalRosso2010}
or in the limit $D=\infty$ \cite{BouchaudMezardParisi1995,LeDoussalMullerWiese2010}. 
In this Letter we present 
a solution for generic $D$, for a non-trivial class of random initial conditions. These appear naturally in related works in the context of elastic interfaces in $D=1$
\cite{ABBM,LeDoussalWiese2008a,LeDoussalWiese2008c,LeDoussalWiese2010a} and more recently in
higher $D$ \cite{LeDoussalWiese2010a}. At this stage 
the solution is a conjecture, based on a field-theory argument.  Here, we state the conjecture and provide an
accurate numerical test.

The decaying Burgers equation describing  a potential flow-velocity field  $\vec v(\vec r,t)=  \vec \nabla \hat V(\vec r,t)$ reads
\bea \label{burg}
\partial_t \vec v = \nu \nabla^2 \vec v -  \frac{1}{2} \vec \nabla \vec v^2\ .
\eea
The inviscid limit corresponds to $\nu=0^+$ \cite{nu}.  Our model is defined by choosing the distribution of the initial velocity field $\vec v(\vec r,0)$ 
as a centered gaussian  with stationary increments $\delta \vec v(\vec r_1,\vec r_2) = \vec v(\vec r_1,0)- \vec v(\vec r_2,0)$  of correlations:
\beq \label{mod}
\frac{1}{2} \overline{\delta v_i(\vec r_0,\vec r_0+\vec r) \delta v_j(\vec r_0,\vec r_0+\vec r) } = \frac{B}{2} |\vec r\,| (\delta_{ij} + \hat r_i \hat r_j)
\eeq
 with statistical translational invariance, where $\hat r=\vec r/|r|$. We denote by  $\overline{\cdots}$ averages over initial conditions. In $D=1$ this reduces to  a Brownian initial velocity
\cite{Sinai1992,SheAurellFrisch1992,Bertoin1998}. 
In general $D$ 
there is no
obvious Markov property, except that the velocity along any given direction is a Brownian. 

We now state our prediction, obtained from field-theoretical considerations, and to be considered a conjecture in the mathematical sense: The characteristic function of the 2-point distribution of velocity increments is
\beq \label{conj}
\overline{ e^{- \vec \lambda \cdot [\vec v(x \vec e_1,t) - \vec v(0,t)] } } = e^{x [Z_t(\vec \lambda)-\lambda_x]} \ .
\eeq
In the notations of equation (\ref{mod}) we set without loss of generality $\vec r_0=\vec 0$, and $\vec r = x \vec e_1$, $x>0$;  $\vec e_1$ is the unit vector along the $x$-axes. The generating function $Z_t(\vec \lambda)$ is  a function of two arguments,  $Z_t(\vec \lambda) = Z_t(\lambda_x,\vec \lambda_\perp^2)$, $\vec \lambda_\perp= \vec \lambda - \lambda_x  \vec e_1$, and will be discussed below. 
 Furthermore,  $Z_t(\vec \lambda)=S_m \tilde Z(S_m \vec \lambda/t)$ where $S_m = B t^2$ is a spatial scale related to a characteristic
size of the shocks defined below.
The function $\tilde Z(\vec \lambda)$ is given below in some special directions, its full expression, not reproduced here, is computed in 
\cite{LeDoussalWiese2010a}. 
The general form (\ref{conj}) can be derived  
under the following assumptions : (i)  velocity increments are localized in shocks;  
(ii) along a given direction, the location of shocks are independent (i.e. Poisson statistics)
(iii) shock sizes are mutually uncorrelated and independent from locations. 
The converse is also true \cite{levy}.
%
%

Both $S_m$ and $\tilde Z(\vec \lambda)$ are related to shocks as follows. 
Let us denote by $x_{\alpha}$ the discrete set of points where the shock manifold intersects the $x$ axis, and (minus) their associated velocity jumps as $\vec S_i= t \big[ \vec v(x_{\alpha}-0^+)-\vec v(x_{\alpha}+0^+) \big]$
where  $\alpha$ labels the shocks.
The shock-size density is $\rho(\vec S)=\overline{\sum_\alpha \delta(x-x_\alpha) \delta(\vec S-\vec S_\alpha)}$. The characteristic
shock size $S_m$ is defined from the moments of the longitudinal shock component $S_x$ as
\beq \label{sm}
S_m = \frac{\langle S_x^2 \rangle}{ 2\langle S_x \rangle}\ ,
\eeq 
where $\langle\cdots \rangle$ denotes averages over the shock density $\rho(\vec S)$ which takes the form
\beq \label{rho}
\rho(\vec S) = \frac1{S_m^2} p\Big(\frac{\vec S}{S_m}\Big) \ .
\eeq 
$p(\vec s)$ is a function of the reduced shock-size $\vec s:=\vec S/S_m$. 
By construction $ \left<s_x^2\right>=2$. The identity 
$\int \rmd \vec S\, \rho(\vec S) S_x=1$ implies a second normalization condition 
$\left< s_x\right>=1$. Note that here and below $\langle \ldots \rangle$ denote moments either over $\rho(S)$ or $p(s)$.  Expanding Eq.\ (\ref{conj}) for small $x$ and using (\ref{rho}) one finds that  $\tilde Z(\vec \lambda)$
is the generating function for the distribution of reduced shock sizes,
\beq
\tilde Z(\vec \lambda) := \left<  e^{\vec \lambda \vec s}-1\right> := \int \rmd \vec s\, \left(e^{\vec \lambda \cdot \vec s} -1\right) p(\vec s)\ .
\eeq
Its expansion has been computed in \cite{LeDoussalWiese2010a}: 
\be \label{series}
\tilde Z(\vec \lambda) = 
 \lambda_{x} +\frac{1}{2} \lambda_{x}^{2}+ \frac{1}{2} \vec \lambda^{2}  + 2 \lambda_{x} \vec
\lambda^{2} + \frac{3}{2} (\vec \lambda^{2})^{2}+\frac{9}{2}\vec
\lambda^{2}\lambda_{x}^{2} -\lambda_{x}^{4} 
+\ldots \ee
It implies universal moment ratios, in particular
\beq \label{rat1}
\frac{\langle S_x^2 \rangle}{\langle S_\perp^2 \rangle} = \frac{2}{\langle s_\perp^2 \rangle} = \frac{2}{D-1} \ .
\eeq
where here  $\vec S_\perp$ and  $\vec s_\perp$ denote the component of the shock orthogonal to $x$.
While the set of shocks along $x$ are uncorrelated 
both in position and size, a property which indeed implies (\ref{conj}), by contrast, longitudinal and transverse components of a given shock are correlated, as from 
(\ref{series}) one can calculate higher moments, e.g.
\beq \label{rat2}
4 \frac{\langle S_x S_\perp^2 \rangle \langle S_x \rangle}{\langle S_x^2 \rangle^2}  
=  \langle s_x s_\perp^2 \rangle = 4 (D-1)\ .
\eeq  
We now indicate the origin of our conjecture, by recalling the connection to disordered systems. Eq. (\ref{burg}) is solved by the 
Cole-Hopf transformation \cite{Burgers74} in the limit $\nu\to 0$:
\bea \label{co}
\hat V(\vec r,t) = \min_{\vec u} \left[ \frac{1}{2 t} (\vec u-\vec r)^2 + V(\vec u) \right]\ ,
\eea
where $V(\vec u)$ is the potential associated with the initial condition, i.e.\ $\vec v(\vec r,t=0)=  \vec \nabla V(\vec r)$. Hence for a random initial condition
the problem is equivalent to finding the minimum energy position of a particle in
a random potential, plus a harmonic well. Denoting by $\vec u(\vec r)$ the position of
the minimum in (\ref{co}), the velocity field is
$\vec v(\vec r,t) = [\vec r-\vec u(\vec r)]/t$. At the shocks, the minimum
jumps, and the shock size is $\vec S=\vec u(x_\alpha + 0^+)-\vec u(x_\alpha - 0^+)$.
Note that $\overline{u(\vec r)}=\vec r$ which implies $\langle S_x \rangle=1$ as stated above. 

The random potential $V(\vec u)$ corresponding to the present model (\ref{mod}) 
is a generalization of the 1D random acceleration process \cite{Burkhardt1993,MajumdarRossoZoia2010} to $D$ dimensions.
To define it one needs a large-scale regularization; we choose 
periodic boundary conditions of period $L$ in all $D$ directions,
\beq \label{H}
V(\vec u)= L^{- \frac{D}{2}} \sum_{\vec q \neq 0} V_{\vec q} e^{i \vec q \cdot \vec u} , \quad 
\overline{V_{\vec q} V_{{\vec q}'}} = \frac{\sigma^2 \delta_{\vec q,-{\vec q}'}}{(q^2)^{\frac{D}{2}+H}} \ ,
\eeq
where $\vec q=\frac{2 \pi}{L} \vec n$, $\vec n \in \left\{-L/2+1,\ldots,L/2-1,L/2  \right\}^D$, in the limit $L \to \infty$, and $H=3/2$.
In real space this leads to a non-analytic cubic potential correlator 
$\overline{V(\vec u) V(\vec u')} = R_0(\vec u-\vec u')$ 
with $R_0(\vec u)-R_0(0)= - \frac{1}{2} A_L u^2 + \frac{B}{6} |u|^3 + O(1/L)$ with 
 $A_L =0.0182 L\sigma^2 +O(L^0)$ 
and $B=\sigma^2/(3 \pi)+O(1/L)$. The initial velocity correlator is
$\overline{v_i(\vec r,t=0) v_j(0,t=0)} = - \partial_i \partial_j R_0(\vec r)$ 
with independent increments distributed as in (\ref{mod}).

\begin{figure}[t]
\includegraphics[width=4.2cm]{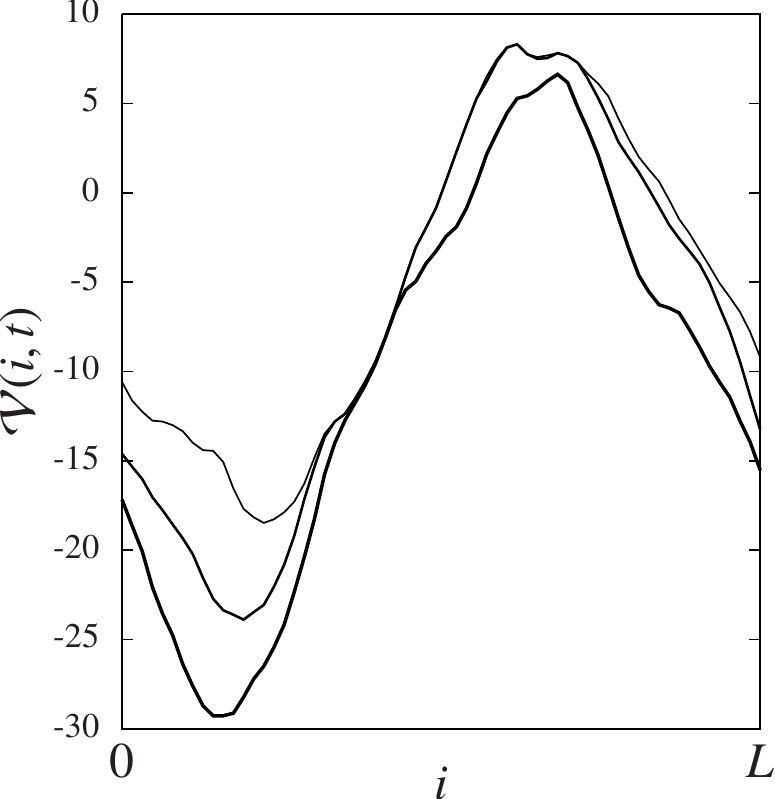} \includegraphics[width=4.2cm]{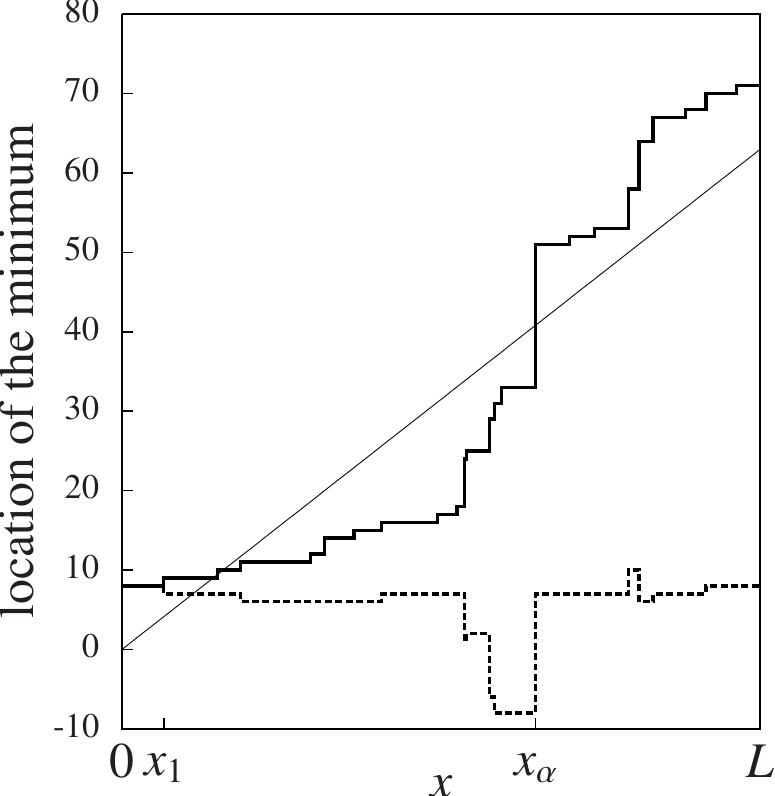}
\caption{\underline{Left.} Step One: {\em reduction to one dimension}. Effective 1-dimensional potential ${\cal V}(i,t)$ after the minimization over $j$, as given in Eq.(\ref{effective}),  at different times (from top to bottom $t=0, 2, 8$). The potential becomes deeper and deeper
as time increases. \underline{Right.} Step Two: {\em location of the minimum}. Solid stair-case line  is $i_{\min}(x)$, dashed line is  $j_{\min}(x)$ for $t=8$. The drift $x$ is indicated. Shocks are only forward in $x$ direction. 
} 
\label{f:1}
\end{figure}

In a nutshell the basis for the conjecture is as follows. The present model is the $d=0$ limit of a model
of an elastic manifold (of internal dimension $d$) in a random potential and a quadratic well of curvature $1/t$. 
The analogous variable to $\vec u(\vec r)$ is the center of mass of the manifold, and $\hat V(\vec r)$ the energy of the optimal configuration as
a function of well position. Its second cumulant defines a renormalized potential disorder
correlator $R(\vec r)$ for any $d$, which is shown to obey a Functional RG equation 
as $t$ is varied. This equation can be solved perturbatively in $R$ in a $d=4-\epsilon$ expansion.
It turns out that the initial correlator $R_0(\vec r)$ corresponding to (\ref{mod}) solves the FRG equation
{\it to all orders} in $\epsilon$, i.e. there are no  loop corrections. 
This implies that the correlation functions need only be computed to tree-level, either by recursion or from a saddle-point method, as detailed in \cite{LeDoussalWiese2010a}. This leads to (\ref{conj}) and to $Z_t(\lambda)$, which hold for any $D$ and any $d$, for this choice of initial conditions,  although we need only $d=0$ (Burgers). 
A further result, proved to lowest order in $\epsilon =4-d$ \cite{LeDoussalWiese2010a}
but which we expect to hold for any $d$, is that (\ref{mod}) is an {\it attractive} fixed point of the RG, hence for velocity correlations which differ from (\ref{mod}) only at small $r$, the behaviour at large $t$
 again follows (\ref{conj}) \cite{nu}. Of course we cannot
exclude non-perturbative corrections, hence our prediction is, strictly, a conjecture. 
In support we note that for $D=1$ it {\it has been proven}
in \cite{Bertoin1998}.  To check it in $D=2$ we now turn to numerics.

A powerful algorithm allows to solve this problem for a slightly  modified version of Eq.~(\ref{co}), with a discretized variable $\vec u=(i,j)$  and a continuous variable $\vec r=x \vec e_1$
\bea \label{codiscrete}
\hat V(x \vec e_1,t) = \min_{1\le i, j \le L}  \left[ \frac{(i-x)^2 }{2 t}   + \frac{j^2}{2 t} + V(i,j) \right],
\eea
for any $x$ in the interval $(0,L)$. Let us now discuss how the algorithm finds the site  $\vec u_{\min}(x)=(i_{\min}(x),j_{\min}(x))$  which satisfies the minimization condition (\ref{codiscrete}):

\underline{Step 1:} {\em Reduction to a 1-dimensional problem.} For each value of $i$  we perform a minimization over the transverse coordinate $j$, keeping in memory the location of the minimum, $j_{\min}^*(i)$. Since this operation does not involve $x$, the effective 
dimension of the problem is reduced to $1$, and Eq.~(\ref{codiscrete}) becomes
\bea \label{codiscrete2}
\hat V(x \vec e_1,t) &=& \min_{1\le i \le L}  \left[ \frac{(i-x)^2 }{2 t}   +{\cal V}(i,t) \right] \ .
\\
 \label{effective}
{\cal V}(i,t) &=& \min_{1\le j\le L}    \left[ \frac{j^2}{2 t} + V(i,j) \right].
\eea
The reduced potential  ${\cal V}(i,t)$ is plotted on Fig.~\ref{f:1} (left).

\underline{Step 2:} {\em Determination of  $i_{\min}(x)$}. The latter is an increasing piecewise constant function of $x$. The minimum location in the original $D=2$ lattice is given by $j_{\min}(x)=j_{\min}^*\big(i_{\min}(x)\big)$.  For $x=0$ the minimum position is  found from Eq.~(\ref{codiscrete2}). Increasing $x$, the minimum remains in $i_{\min}(x=0)$ up to a threshold $x_{1}$, above which the minimum takes a new value $i_{\min}(x_{1})> i_{\min}(0)$.  For all $i>i_{\min}(x=0)$ we find the value of $x$ satisfying
\beq \label{codiscrete3}
   \frac{(i-x)^2 }{2 t}   +{\cal V}(i,t) =\frac{(i_{\min}(0)-x)^2 }{2 t}   +{\cal V}(i_{\min}(0),t)\ .
\eeq
$x_{1}$ is the smallest value of $x$ for which this condition is satisfied. One then searches the next minimum 
\begin{figure}
\includegraphics[width=4.2cm]{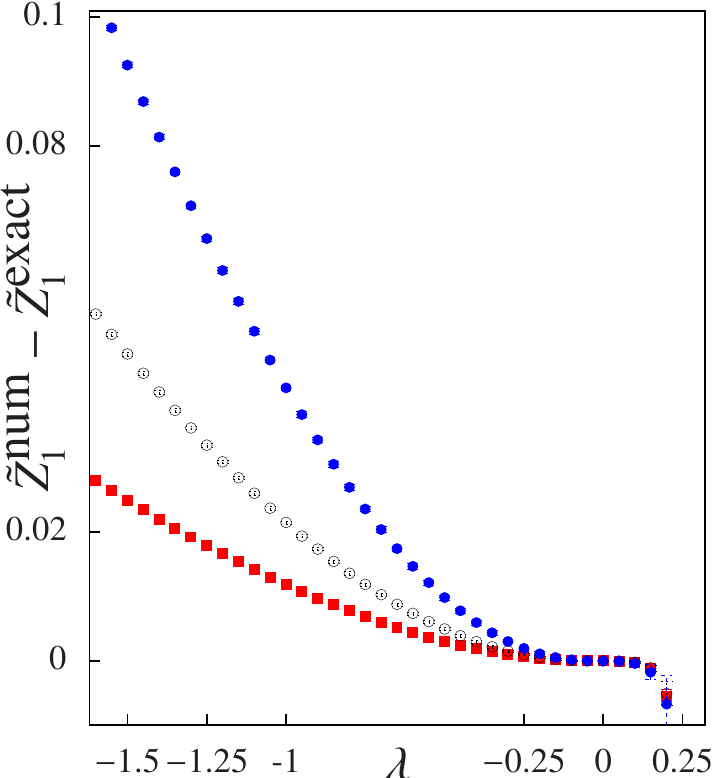} \includegraphics[width=4.2cm]{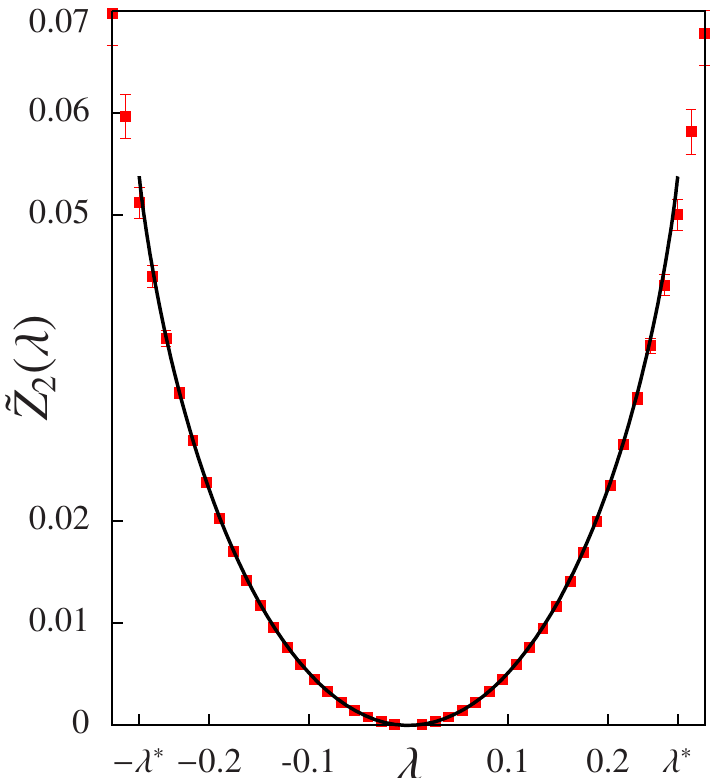}
\caption{Color Online.  \underline{Left}: Convergence of the measured $ \tilde Z^{{\rm num}}_1(\lambda)$ to the analytical prediction 
 (\ref{pred1}) for different times: $t=2.04$ (filled circles), $t=2.78$ (open circles), $t=4$ (squares). 
 \underline{Right}:  measured $ \tilde Z_2(\lambda)$  for $t=4$ (squares) compared to the prediction  (\ref{z2}) (solid line).  \label{Z1}}
\end{figure}%
and the procedure is iterated up to $x=L$, see Fig.~1 (right). 

\underline{Step 3:} {\em Shock sizes}. 
Given  the sequence of minima locations $\vec u_{\min}(x)=(i_{\min}(x),j_{\min}(x))$, the shocks sizes $\vec S$ are the discontinuities in these piecewise functions of $x$. The velocity profile is $v_x(x,t) = ( x-i_{\min}(x))/t$, $v_y(x,t) = -j_{\min}(x)/t$.  Note that in our discrete model the shock size is cut off from below at $S_0=1$ and from above at $L$. Self-affine scaling and Eq.~(\ref{burg}) are expected to hold in the  continuum limit when  
$S_0 \ll S_m  \ll L$ or equivalently $S_0/S_m=1/(B t^2)\ll s \ll L /(B t^2) $.

\begin{figure} \label{t}
\includegraphics[width=4.2cm]{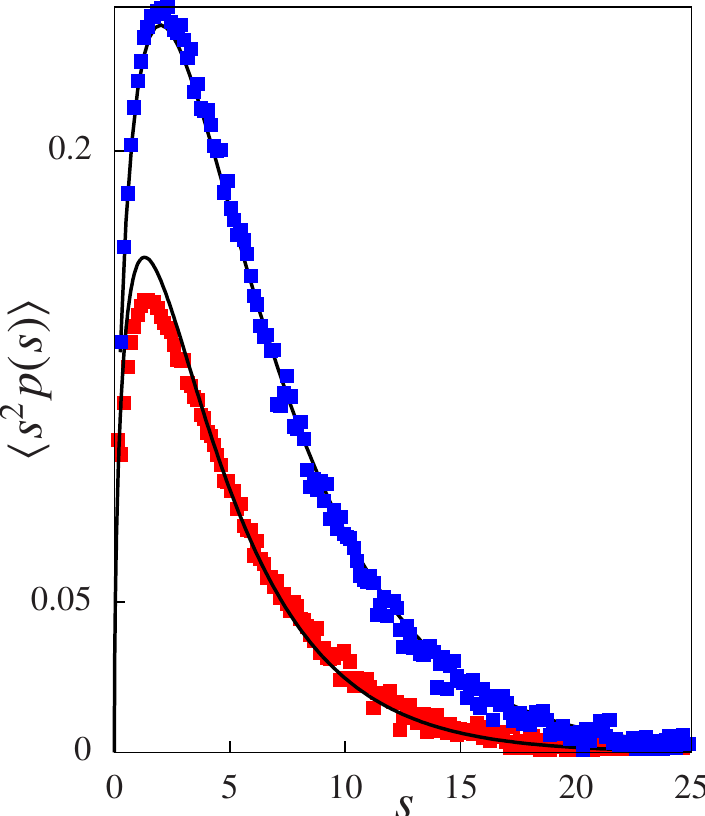} \includegraphics[width=4.2cm]{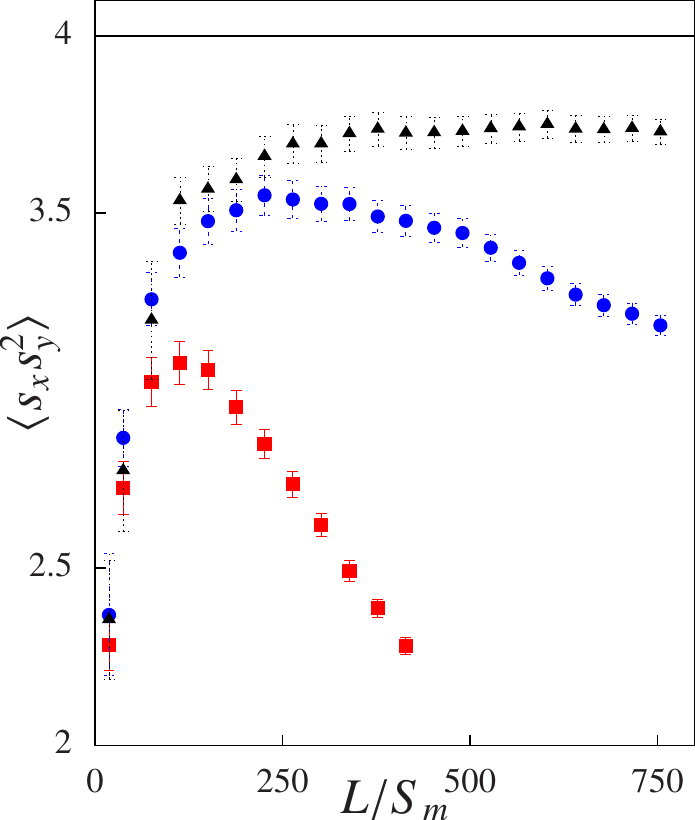}
\caption{Color Online. \underline{Left}.  Plot of $s_x^2 p_1(s_x)$ (top curve) and $s_\perp^2 [p_2(s_\perp)+ p_2(-s_\perp)]$ (bottom curve).  
Shocks at $t=8$ ($S_m=7\pm 0.1$). Solid line represents the analytical predictions.
 \underline{Right}. Test of the ratio (\ref{rat2}).
Squares are for $L=2^8$, circles  for $L=2^{10}$, and triangles for $L=2^{12}$. Numerical data approach the predicted value when $L$ is large.  For  $L/S_m \to 0$ the particle feels the periodic potential, i.e.\ $S_x=L$ and $S_\perp=0$.  For  $t \to 0$,  lattice spacing is important: $S_x=1$ and $S_\perp=0$. The plateau value
is consistent with $4-c_1 L^{-1/2}$. \label{sxsy}}
\end{figure}

(iv) \underline{Step 4:} {\em Numerical implementation}. 
In practice, we consider  a $D=2$ square latice (usually of size $L=2^{12}$), the correlated random potential $V(i,j)$ is constructed from $L^2$ independently distributed Gaussian random numbers via a ``fast Fourier transform''
of Eq.\ (\ref{H}).  Note that the sum over the components of $\vec n$ are now running over integers from $-L/2+1$ to $L/2$.
 The zero mode
$\vec n=0$ is set to zero, $V_0=0$. We choose $\sigma^2=1$, which implies that $B=1/(3 \pi)$ in formula (\ref{mod}). 
We collected a large number of shocks ($\sim 10^6 - 10^7$) using many samples, from which we computed $S_m$ and verified
the prediction $S_m = t^2/(3 \pi)$. From the reduced sizes $\vec s_\alpha:=\vec S_{\alpha}/S_m$,  we measured $\tilde Z(\lambda_x,\lambda_\perp) = \frac{1}{N} \sum_\alpha (e^{\vec \lambda \cdot \vec s_\alpha} -1 )$, specifically
$\tilde Z_1(\lambda):=\tilde Z(\lambda,0)$ and $\tilde Z_2(\lambda):=\tilde Z(0,\lambda)$.
The conjecture states that for the longitudinal component of the shock
\beq  \label{pred1}
Z_1(\lambda)=\frac{1}{2}(1-\sqrt{1-4 \lambda})\ , \quad  p_1(s) = \frac{1}{2 \sqrt{\pi} s^{3/2}} e^{-s/4}
\eeq
with $p_1(s_x):=\int \rmd s_\perp p(s_x,s_\perp)$, i.e. the same as obtained for $D=1$ \cite{Bertoin1998,Valageas2009}
and for the related Galton process \cite{WatsonGalton1875,LeDoussalWiese2008a}. This is verified on Fig.~\ref{Z1} (left) and
Fig.~\ref{sxsy} (left). Since the agreement is very good, we have plotted on Fig.~\ref{Z1} (left) the difference with the analytical prediction
to emphasize the small deviations.  These deviations are more important for large negative
 $\lambda \sim - 1/s_0$, sensitive to the small lattice cutoff $s_0=S_0/S_m$ for the reduced shock sizes. Increasing the
time, $s_0$ decreases as $s_0 \sim 1/t^2$. 
The prediction for the characteristic function of the $y$ component of the shock
sizes, $\tilde Z_2(\lambda)$, is obtained by elimination of $\theta$ in the system of equations
\begin{eqnarray} \label{z2}
 \lambda (\theta) &=&  \sin \theta \frac{ \sqrt{5-\cos (4 \theta)}+2}{\big[1-\cos (2 \theta)+\sqrt{5-\cos (4
   \theta)}\big]^2}  \qquad \\
\tilde Z_2(\theta) &=& \frac{\cos \theta}{2} \frac{ \sqrt{5-\cos (4 \theta)}-2}{1-\cos (2 \theta)+\sqrt{5-\cos (4
   \theta)}} \ .
\end{eqnarray}
Numerically, the Laplace inversion can be performed  
 to determine $p_2(s_\perp)=\int \rmd s_x\, p(s_x,s_\perp)$ with high precision. (It is an integral over a segment of $\theta$ in the complex plane.) $p_2(s)$ is plotted on Fig.\ \ref{sxsy} (left). For large $s$,  $p_2(s) \approx 1.7304 |s|^{-5/2} e^ {-0.2698 |s|} $; while for small $s$ $p_2(|s|) = 0.12375|s|^{-3/2}$. We have plotted the measured and calculated $\tilde Z_2(\lambda)$ on Fig.~\ref{Z1} (right).  Since $p_2(s)$ is symmetric in $s$, the same holds true for  $Z_2(\lambda)$.  The left and right edges of the analytic curve  are at $|\lambda^*|=0.2698...$, the constant in the exponential decay of $p_2(s)$. The agreement is excellent up to this point, where the size $L$ cuts the divergence for $|\lambda| > \lambda^*$.

\begin{figure}[t]
\includegraphics[width=7.7cm]{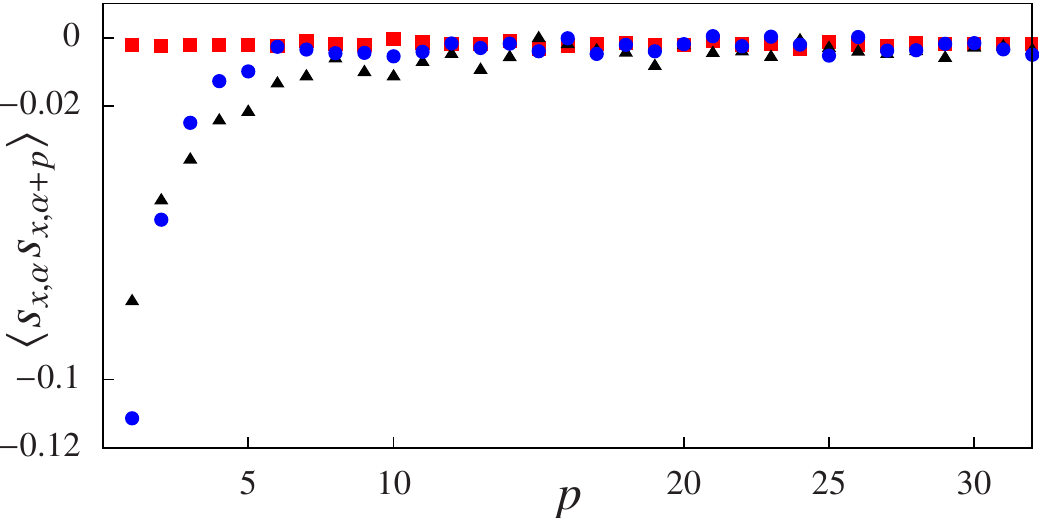} \includegraphics[width=7.5cm]{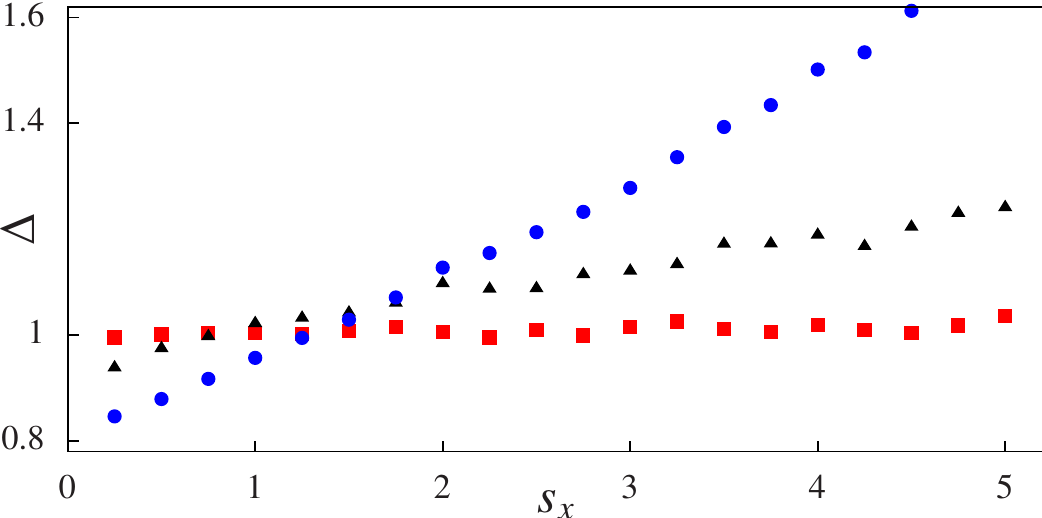}
\caption{Color Online. Circles corresponds to $H=1/2$ , Triangles to $H=1$ and Squares to $H=3/2$.
\underline{Top:} Connected size correlations of subsequent shocks. Correlation decay is slower for $H=1$,
for $H=3/2$  no correlation has been detected.
\underline{Bottom:} Normalized shock distance. 
} 
\label{last}
\end{figure}

We now discuss shock correlations:  {\em First}, the universal ratio  (\ref{rat1}) was measured to be $2.034 \pm 0.015$, very close to its analytical prediction. 
{\em Second}, correlations of jumps in the different directions are measured by the ratio (\ref{rat2}), plotted on Fig.~\ref{sxsy} (right).  In both cases, the deviations can be attributed to  finite-size corrections, see the caption of Fig.\ 3. {\em Third}, we studied the correlations between {\it subsequent
shocks}. To emphasize the remarkable nature of the present model ($H=3/2$) we compare with two other ones, 
with potential given by (\ref{H}) with $H=0.5$ and $H=1$. In Fig.~\ref{last} (top) we show the connected correlation
$\langle s_{x,\alpha} s_{x,\alpha+p} \rangle^c$ of the longitudinal size $s_{x,\alpha}$ of a shock with the $p$-th subsequent shock. Fig.~\ref{last} (bottom) focuses on  the  correlations between the location and the size of shocks.
For example, large shocks are more  isolated with respect to small shocks? To check this we compute the average distance 
$\langle x_{\alpha+1}-x_\alpha \rangle$ between consecutive shocks, normalized by its averaged and called $\Delta(s_x)$,
as a function of the longitudinal size $s_{x}$ of the shock $\alpha$. Fig.~\ref{last} (bottom) shows strong correlations for $H=0.5$ and $H=1$.  For $H=3/2$ no effect was detected. This remarkable statistical independence of the shocks is essential for the main formula (\ref{conj}).

To conclude, we proposed a conjecture for the $D$-dimensional decaying Burgers equation with initial conditions
which generalize the Brownian  for $D=1$. We tested it numerically and checked that the
shocks and the velocity increments along an axis are statistically independent at any time $t$.
The conjecture is based on vanishing loop corrections in the field theory for the disordered problem and
 its generalization to elastic manifolds. Although confirmed within our numerical accuracy, any deviation 
 would have important consequences for the -- probably non-perturbative --
corrections to the field theory. We hope this motivates efforts to prove or infirm our conjecture
on a rigorous basis. 

 This work was supported by ANR grant 09-BLAN-0097-01/2 and in part
by  NSF grant PHY05-51164. We thank the 
KITP for hospitality.




\begin{thebibliography}{10}

\bibitem{GawedzkiKupiainen1995}
K.~Gaw\c edzki and A.~Kupiainen,
\newblock Phys. Rev. Lett. {\bf 75} (1995)   3834.
L.T. Adzhemyan, N.V. Antonov  and A.N. Vasil'ev,
\newblock Phys. Rev. {\bf E 58} (1998)   1823--1835.
K.J. Wiese,
\newblock J. Stat. Phys. {\bf 101} (2000)   843--891.



\bibitem{Burgers74}
J.M. Burgers,
{\em The non-linear diffusion equation,}
\newblock Dordrecht, 1974.
J.\ Bec and K.\ Khanin,
\newblock Phys. Rep. {\bf 447} (2007)   1--66.
U.~Frisch and J.~Bec,
\newblock in {\em New Trends in Turbulence,} Springer EDP-Sciences, 2001.

\bibitem{Kida1979}
S.~Kida,
\newblock J. Fluid. Mech. {\bf 93} (1979)   337.

\bibitem{Sinai1992}
Ya~G. Sinai,
\newblock Commun. Math. Phys. {\bf 148} (1992)   601.

\bibitem{SheAurellFrisch1992}
Z.-S. She, E.~Aurell  and U.~Frisch,
\newblock Commun. Math. Phys. {\bf 148} (1992)   623.

\bibitem{Bertoin1998}
J.~Bertoin,
\newblock Commun. Math. Phys. {\bf 193} (1998)   397--406.

\bibitem{FrachebourgMartin2000}
L.~Frachebourg and P.A. Martin,
\newblock J. Fluid. Mech. {\bf 417} (2000)   323.

\bibitem{LeDoussal2006b}
P.~{Le Doussal,}
\newblock Europhys. Lett. {\bf 76} (2006)   457--463.%

\bibitem{LeDoussal2008}
P.~{Le Doussal,}
\newblock Annals of Physics {\bf 325} (2009)   49--150.

\bibitem{Valageas2009}
P.~Valageas,
\newblock J. Stat. Phys. {\bf 137} (2009)   729.

\bibitem{FyodorovLeDoussalRosso2010}
Y.V. Fyodorov, P.~Le Doussal  and A. Rosso,
\newblock EPL {\bf 90} (2010)   60004.

\bibitem{BouchaudMezardParisi1995}
JP. Bouchaud, M.~M\'ezard  and G.~Parisi,
\newblock Phys. Rev. E {\bf 52} (1995)   3656--3674.

\bibitem{LeDoussalMullerWiese2010}
P.~Le Doussal, M.~M\"uller  and K.J. Wiese,
\newblock EPL {\bf 91} (2010)   57004.



\bibitem{LeDoussalWiese2008c}
P.~{Le~Doussal} and K.J. Wiese,
\newblock Phys. Rev. E {\bf 79} (2009)   051106.

\bibitem{ABBM}
B.~Alessandro, et al. 
\newblock J. Appl. Phys. {\bf 68} (1990)   2901.

\bibitem{LeDoussalWiese2008a}
P.~Le Doussal and K.J. Wiese,
\newblock Phys. Rev. E {\bf 79} (2009)   051105.

\bibitem{LeDoussalWiese2010a}
P.~{Le Doussal} and K.J. Wiese,
to be published.



\bibitem{nu}
 A small $\nu>0$ gives a small width to shocks, which
here scales to zero in reduced units, and is thus irrelevant.

\bibitem{levy}
These are called Levy processes in probability theory. In $D=1$, see J.\ Bertoin, {\em Levy processes}, Cambridge University
Press, 1996.

\bibitem{Burkhardt1993}
T.W. Burkhardt,
\newblock J. Phys. A {\bf 26} (1993)   L1157.

\bibitem{MajumdarRossoZoia2010}
S.~N. Majumdar, A.~Rosso  and A.~Zoia,
\newblock J. Phys. A {\bf 43} (2010)   115001.

\bibitem{WatsonGalton1875}
H.W. Watson and F.~Galton,
\newblock Journal of the Anthropological Institute of Great Britain {\bf 4}
  (1875)   138--144.

\end{thebibliography}

\vspace{100cm}

\end{document}